\documentclass[superscriptaddress,aps,pre,floatfix,twocolumn,twopage,notitlepage,headsepline]{revtex4}
\usepackage{fancyhdr}
\usepackage{amsmath,amssymb}

\usepackage{graphicx,psfrag}% Include figure files \usepackage{bm}% bold math
\usepackage{multirow}
\newcommand{\nn}{\nonumber\\} 
\newcommand{\ti}[1]{^{(#1)}}

\begin{document} 
\title{Cooperative searching for stochastic targets}
\author{Vadas Gintautas} 
\email{vgintautas@chatham.edu}
\affiliation{Center for Nonlinear Studies, Los Alamos National Laboratory, Los Alamos NM 87545, USA} 
\affiliation{Applied Mathematics and Plasma Physics, Theoretical Division, Los Alamos National Laboratory, Los Alamos NM 87545, USA}
\altaffiliation{Physics Department, Chatham University, Pittsburgh PA 15232, USA}
\author{Aric Hagberg} %\email{hagberg@lanl.gov} 
\affiliation{Applied Mathematics and Plasma Physics, Theoretical Division, Los Alamos National Laboratory, Los Alamos NM 87545, USA}
\author{Lu{\'i}s M.
A. Bettencourt} %\email{lmbett@lanl.gov} 
\affiliation{Applied Mathematics and Plasma Physics, Theoretical Division, Los Alamos National Laboratory, Los Alamos NM 87545, USA}
\affiliation{Santa Fe Institute, 1399 Hyde Park Road, Santa Fe NM 87501, USA}

\date{\today}

\begin{abstract}
Spatial search problems abound in the real world, from locating hidden nuclear or chemical sources to finding skiers after an avalanche. We exemplify the formalism and solution for spatial searches involving two agents that may or may not choose to share information during a search. For certain classes of tasks, sharing information between multiple searchers makes cooperative searching advantageous. In some examples, agents are able to realize synergy by aggregating information and moving based on local judgments about maximal information gathering expectations. We also explore one- and two-dimensional simplified situations analytically and numerically to provide a framework for analyzing more complex problems. These general considerations provide a guide for designing optimal algorithms for real-world search problems.
\end{abstract} 

\maketitle
\section{Introduction} 

In the real world, there are many spatial search problems that involve multiple agents or searchers. Communication between such agents or between agents and a centralized command center may be sensitive, costly, or difficult for various reasons. In this work, we explore spatial search problems in this context and examine the classes of search problems for which communication and coordination between multiple agents will enable quantitative advantages over independent information gathering. These problems are very general and can be formalized and solved in terms of information theory. The solutions are essential to the development of quantitative decision support tools under uncertainty (e.g. in medicine~\cite{Fox_2001}) and for
automated multi-agent searches for stochastic 
information~\cite{Agogino_2008}, such as those involving 
distributed sensor networks~\cite{Zhang_2002}.

A search may be thought of as a series of steps by which an agent or agents reduce the uncertainty of the location of a target to zero. This location may be in physical space or in a ``space of possibilities,'' i.e., a set of alternative scenarios. For example, if you need to find your keys in the morning before going to work and there are five rooms in your house, initially you know your keys must be in one of these rooms. After thoroughly searching one room and not finding the keys, your uncertainty is reduced; the keys must now be in one of the four remaining
rooms, and so on. If you find the keys in the second room, then the uncertainty of their location is immediately zero because you know with complete certainty that they are in your hand.

Now imagine you have a friend help you search. The two of you would be searching twice as fast since, assuming you and your friend share information, a pair of searchers can eliminate rooms at a rate twice as fast as that of a single searcher. In more complex search problems, the target might emit some kind of complicated signal that makes it possible for multiple coordinated searchers to dramatically increase efficiency (imagine there are one thousand rooms to search but the keys are attached to some kind of radioactive homing beacon or can emit a sound). In these situations, understanding the nature of the clues and sharing information can be extraordinarily beneficial.

A search whereby agents move based {\em expressly} on information cues rather than following gradients is known as an infotaxis search. In a 2007 paper, Vergassola et al. explored infotaxis in the context of a moth following a pheromone trail through air to find a mate. The moth was performing a spatial search in turbulent air currents that carried the trail h~\cite{Vergassola_2007}.  
Compared to more conventional methods such as chemotaxis (following a chemical
gradient~\cite{Eisenbach_2004, Russell_2003}), infotaxis gives an advantage in  
situations when the signals from the target of the search are
uncertain.  
This might be the case if signals from the target are stochastic, difficult to measure, or highly varying in time. In the case of the moth, the signal was sparse and widely dispersed by the turbulent air currents. When the moth followed the gradient of the strength of the trail directly – chemotaxis – it was forced to take a very circuitous route due to the turbulent dispersal of the trail. However, when the moth employed an infotaxis algorithm to move to positions where it might gain the most information about the source of the trail, its performance improved significantly. Since we will consider only search problems such as these, in the following we will refer to the target of an infotaxis search as the ``source.'' The formalism of infotaxis balances the competing goals of exploiting the current information available and exploring to gain more information, a familiar compromise from other unsupervised learning methods, such as reinforcement learning~\cite{Sulton_1998}.  While effective path planning 
algorithms may be based on the optimization of some objective function, these often
rely on exploiting some features of the known environment~\cite{Hollinger_2009} rather than
a solid foundation based on information theory.  Infotaxis provides a framework for studying
search problems in general and is therefore broadly applicable.

In this article, we explore conditions in which cooperation between multiple infotaxis agents is advantageous. We focus on examples in which agents are able to realize synergetic cooperation by aggregating information and moving based on a local infotaxis algorithm. {\em Synergy}, and its opposite, {\em redundancy}, are information theoretic quantities that are defined in terms of relative probabilities of the stochastic variables involved~\cite{Bettencourt_2008}.  In a recent paper, 
we showed that spatiotemporal correlations are necessary for
synergy~\cite{Gintautas_SBP09}.  When synergy is exploited effectively
it can lead to an exponential reduction in the search effort, in terms of
time, energy, or number of steps~\cite{Seung_1992,Freund_1997,Fine_2002}.  
Here we use a simple one-dimensional search
example and a more realistic two-dimensional generalization 
to show how correlations lead to synergy.
These simple examples provide a framework for analyzing more complex
problems.  Since, in general, the computational cost is greater for
searchers to communicate and perform coordinated movements instead of
moving based on independent decisions, we will describe situations in
which coordination is worthwhile.

\section{Information theory approach to stochastic search} 
Effective and robust search methods for locating stochastic sources
balance the competing strategies of exploration and
exploitation~\cite{Sulton_1998}.  Given a current estimated
probability distribution for the location of a source a searcher might
either exploit the data already collected by moving towards the
location that maximizes this likelihood or sharpen the distribution
(reduce uncertainty further) by moving to gather more diverse data.
The infotaxis search balances these two strategies by optimizing the
expected information gain over the possible next search moves.  In the
following we review some basic concepts from information theory and
formalize the infotaxis algorithm in terms of these quantities.

%In Section \ref{sec:synergy_and_redundancy} we summarize the definitions
%from information theory relevant to the formalism of infotaxis,
%and in Section \ref{sec:spatial_infotaxis} we review the basic algorithm.

\subsection{Information, synergy, and redundancy}  
\label{sec:synergy_and_redundancy}
To determine 
whether searchers can be effectively coordinated we define
define synergy and redundancy as information theoretical
quantities~\cite{Cover_1991} and use them as a measure of coordination.
Synergy is found
when measuring two or more variables {\em together} with respect to another
(e.g. the source's signal) results in a greater information gain than the
sum of that from each variable {\em separately}~\cite{Bettencourt_2007,
Bettencourt_2008}. 
In search problems, synergy is advantageous because then the coordination of two or more searchers is more efficient than the same searchers working independently. In this section we will introduce these concepts in general terms before applying them to a specific search problem.

Consider the stochastic
variables $X_{i}, i=1\ldots n$.  Each variable $X_i$, 
representing a  searcher or source location,
can take on specific
states, denoted by the corresponding lowercase letter $x_i$.
For a single variable $X_i$ the Shannon entropy
(henceforth ``entropy'') is 
\begin{equation*}
S(X_i) = -\sum_{x_i} P(x_i)\log_{2} P(x_i),
\end{equation*}
where $P(x_i)$
is the probability that the variable $X_i$ takes on the value
$x_i$~\cite{Cover_1991}.  
The sum is over all of the possible states $x_{i}$; since $P(x_{i})<1$ always,the entropy is always positive. The entropy is a measure of uncertainty about the
state of $X_i$, therefore entropy can only decrease or remain unchanged as more
variables are measured.  The conditional entropy of a variable $X_1$ given a
second variable $X_2$ is 
\begin{equation*}
    S(X_{1}|X_{2}) = -\sum_{x_{1}, x_{2}} P(x_{1},x_{2})\log_{2} \frac{P(x_{1},x_{2})}{P(x_{2})}\leq S(X_{1}).
\end{equation*}
This expression contains a sum over the joint probability distribution of two variables. Since measuring a second variable can only decrease entropy (or leave it unchanged), the conditional entropy is bounded above by the entropy of the first variable. The mutual information between two variables, which plays an important role in search strategy, is defined as the change in entropy when a variable is measured:
\begin{equation*}
I(X_{1},X_{2}) = S(X_{1}) -  S(X_{1}|X_{2}) \geq 0.
\end{equation*}

This is also the difference between the entropy of one variable and its entropy conditioned on the measurement of a second variable. Mutual information is always positive. These definitions can be directly extended to multiple variables. Just as entropy may be conditioned on an additional measurement, mutual information may be conditioned on the knowledge of other variables. These quantities may be used to generate new information theoretic constructs that we will use in specific search problems.  For three 
variables~\cite{Schneidman_2003} the quantity
\begin{equation}
R(X_{1},X_{2},X_{3}) \equiv I(X_{1},X_{2}) - I(\{X_{1},X_{2}\}|X_{3})
\label{eq:Rdefn}
\end{equation}
 measures the degree
of ``overlap'' in the information contained in variables $X_{1}$ and $X_{2}$
with respect to $X_{3}$.  The sign of this quantity is meaningful.  If $R(X_{1},X_{2},X_{3}) > 0$, there is overlap and
$X_{1}$ and $X_{2}$ are said to be redundant with respect to $X_{3}$.  If
$R(X_{1},X_{2},X_{3}) < 0$, more information is available when these variables
are considered together than when considered separately.  In this case $X_{1}$
and $X_{2}$ are said to be synergetic with respect to $X_{3}$.  If
$R(X_{1},X_{2},X_{3}) = 0$, $X_{1}$ and $X_{2}$ are
independent.

\subsection{Bayesian inference and spatial infotaxis} \label{sec:spatial_infotaxis}

We first formulate the general spatial stochastic search problem for
$N$ searchers seeking to find a stochastic source located in a finite,
$D$-dimensional space.  This is a generalization of the single
searcher formalism presented in Ref.~\cite{Vergassola_2007}.  At any
time step, the searchers $s_i$, $i=1,2,\ldots,N$, are located at
position $r_{i}$ and observe some number of particles $h_{i}$ from
the source.  The searchers do not get information about the
trajectories or speed of the particles; they only get information if a
particle was observed or not.  Therefore simple geometrical methods
such as triangulation are not possible.

Consider a random variable $R_0$, which can assume a number of specific values, denoted by $r_0$.  The values of $r_0$ refer to positions in space that may contain the stochastic source.  
Only one value of $r_{0}$ corresponds to the (yet unknown) location of
the source $s_0$.  The searchers compute and share a probability
distribution $P\ti{t}(r_{0})$ for the source location at each time
index $t$.  Initially the probability for the source $P\ti{0}(r_{0})$
is assumed to be to be uniform.  After each measurement
$\{h_{i},r_{i}\}$, the searchers update their estimated probability
distribution of source positions via Bayesian
inference~\cite{Bernardo_1994} and decide what move to make 
(possibly remaining at the same position).
The goal of Bayesian inference is to improve an estimated probability distribution $P(X)$, where $X$ is a random variable that can assume a set of values denoted by ${x}$. Assuming that $Y$ is another random variable (that can assume a set of values denoted by ${y}$) and that $X$ and $Y$ are not independent (that is, $I(X,Y)\neq 0$), knowledge of the state of $Y$ can be used to improve $P(X)$. After a measurement reveals $Y=y$, the probability of this measurement given the current estimated	is computed. The probability of this measurement is $P(Y=y|X)$. Bayesian inference makes it possible to assimilate this information into the current estimate of $P(X)$ via a Bayesian update step: $P(X)=P(X|Y=y)=P(Y=y|X)P(X)/A$, where $A$
is a normalization factor. This step includes an explicit statement of equivalence because each new measurement is included implicitly in $P(X)$. Therefore the measurement $Y=y$ improves the estimate of $P(X)$. The searchers will use this Bayesian inference framework to improve their estimate of the probability distribution of source locations $P^{(0)}(r_{0})$ after each measurement ${h_{i},r_{i}}$.

To decide where to move next, the searchers follow an infotaxis algorithm
for multiple searchers.  First the conditional probability 
\begin{equation}
    P\ti{t+1}(r_{0}|\{h_{i},r_{i}\}) \equiv 
    \frac{1}{A}P\ti{t}(r_{0})P(\{h_{i},r_{i}\}|r_{0})\,,
\end{equation}
is calculated, where $A$ is a normalization
over all possible source locations $r_0$ as required by Bayesian inference.  This is then assimilated via Bayesian update,
\begin{equation}
P\ti{t+1}(r_{0}) \equiv P\ti{t+1}(r_{0}|\{h_{i},r_{i}\}).
\end{equation}
If the searchers do not find the source at their present
locations they choose the next local move using an infotaxis step
to maximize the expected information gain.
The expected information gain is computed in the following way.
The entropy of the distribution $P\ti{t}(r_{0})$ at time $t$ is
defined as
\begin{equation}
S\ti{t}(R_{0}) \equiv -\sum_{r_{0}} P\ti{t}(r_{0}) \log_{2} P\ti{t}(r_{0}).
\end{equation}
For a specific measurement $\{h_{i},r_{i}\}$
the entropy {\em before} the Bayesian update is 
\begin{equation}
S\ti{t}_{\{h_{i},r_{i}\}}(R_{0}) \equiv 
-\sum_{r_{0}}P\ti{t}(r_{0}|\{h_{i},r_{i}\})\log_{2} P\ti{t}(r_{0}|\{h_{i},r_{i}\}).
\end{equation}
We define the difference between the entropy at time $t$ and the 
entropy at time $t+1$ after a measurement $\{h_{i},r_{i}\}$ to be 
\begin{equation}
\Delta S\ti{t+1}_{\{h_{i},r_{i}\}} \equiv 
S\ti{t+1}_{\{h_{i},r_{i}\}}(R_{0}) - S\ti{t}(R_{0}).
\end{equation}
For a uniform prior, $P\ti{0}(r_0)=1/M$ for $M$ possible locations of
the source in the discrete space, the entropy is maximum,
 $S\ti{0}(R_{0})=\log_{2} M$.
For each possible joint move $\{r_{i}\}$, the change in
expected entropy $\overline{\Delta S}$ is computed and the move with the
minimum (most negative) $\overline{\Delta S}$ is executed.

The expected information gain is found by computing the entropy change
for all of the possible joint searcher moves
\begin{align}
    \overline{\Delta S} &= -\biggl[\sum_{i}P\ti{t}(R_{0}=r_{i})\biggr] S\ti{t}(R_{0})
    + \biggl[1 - \sum_{i}P\ti{t}(R_{0}=r_{i})\biggr]\nn
    &\times \sum_{ h_{1}, h_{2} } \Delta S\ti{t+1}_{\{h_{i},r_{i}\}}
     \biggl[\sum_{r_{0}} P\ti{t}(r_{0})P\ti{t+1}(\{h_{i},r_{i}\}|r_{0})\biggr].
\label{eq:deltasbar}
\end{align}
The first term in Eq.~\eqref{eq:deltasbar} corresponds to one of the
searchers finding the source in the next time step (the final entropy will be $S=0$ so $\overline{\Delta S}=-S$).
The second term considers the reduction in entropy for all possible measurements
at the proposed location, weighted by the probability of each of those
measurements.  The probability of the searchers obtaining the
measurement $\{h_{i}\}$ at the location $\{r_{i}\}$ is given by the
trace of the probability $P\ti{t+1}(\{h_{i},r_{i}\}|r_{0})$ over all
possible source locations.

At each step the searchers move jointly to increase the expected information
gain as measured by the change in entropy of the probability distribution.
%Thereby the searchers move to reduce the uncertainty in the location of the source to zero as quickly as possible.  
Although this algorithm is general in the following we consider only
the case of two searchers ($N=2)$ and both one- and two-dimensional
spatial domains.

\section{Searching for correlated signals in one dimension} 

\label{sec:angular}

Sources that emit uncorrelated signals provide no opportunity 
for coordination because the searchers are never 
synergetic~\cite{Gintautas_SBP09}.
We instead consider signals with spatial, temporal, or other
correlations. The simplest nontrivial example is searching in a
one-dimensional domain for a source that emits two particles
simultaneously in opposite directions.  Two searchers should be able
to exploit the correlations in the signal; if both searchers
simultaneously observe particles, they can immediately conclude that
the source is located between them.  Therefore we expect synergy to be
possible for some spatial arrangements of the source and searchers.

First consider a finite one-dimensional domain with a source $s_0$ and
two searchers $s_1$ and $s_2$ at the corresponding positions $\{r_0,
r_1, r_2\} \in [0,1]$.  The source is assumed emit two particles
simultaneously and in opposite directions.  That is, one particle is
emitted to the left and one to the right of the source.  The two
searchers $s_1$ and $s_2$ are identical with a fixed cross section
such that $0<a<1$ is the probability of a searcher capturing one of
the particles emitted from the source.  At each step in the search
the number of particle ``hits'' measured by searchers $s_1$ and $s_2$
are denoted by $h_{1}\in\{0,1\}$ and $h_{2}\in\{0,1\}$, respectively.

\begin{figure}[htb] 
    \includegraphics[width=1.0\linewidth]{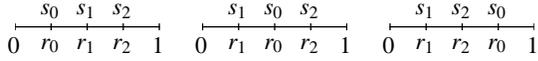} 
    \caption{The three unique cases for relative positions of
      the source and two identical searchers on a one-dimensional
      domain.  Since the probably of detection
      in this example does not depend on distance, we need only
      consider the spatial arrangement of the source and searchers.
      The cases are labeled  $s_is_js_k$ according to the relative
      spatial ordering for the source $s_0$ and searchers $s_1, s_2$.
      Since the searchers are identical, $s_1$ and $s_2$ are 
      interchangeable and there are only three unique cases.} 
    \label{fig:1d_diagram}
\end{figure}

To calculate $R(r_0,h_1,h_2)$ it is first necessary to compute
the probabilities of $h_i$ for each searcher given the position
of the source $r_0$. We note that since there is no distance 
dependence in the capture probability $a$, it is sufficient to 
consider three separate cases depending on the relative
positions, or ordering, of the source and the searchers, as shown in 
Fig.~\ref{fig:1d_diagram}.  For example, 
if $s_1$ is to the left of the source and
$s_2$ is to the right of the source, the order is $s_1s_0s_2$.
Note this is equivalent to the case
$s_2s_0s_1$  since the searchers are identical.

% For this problem setup the random variables are the source position
% $r_{0}$, and the number of particles $h_{1}$, and $h_{2}$ observed by
% $s_1$ and $s_2$, respectively.  The positions $r_{1}$ and $r_{2}$, are
% not random variables.  Instead every quantity (such as a probability
% distribution) iscomputed as a function of $r_1$ and $r_2$.

If a searcher observes a particle, it is assumed to be absorbed so that the other searcher will not be
able to observe it.  For example if $s_1$ is between $s_2$
and the source (case $ s_0s_1s_2$ or $ s_2s_1s_0$) then the probability that
$s_2$
observes a particle depends on whether $s_1$ observed it.  If $s_1$
observed it, then the probability that $s_2$ observes it must be 0.  If
$s_1$ did not observe it then the probability that $s_2$ will observe
it is $a$
\begin{align}
&P(h_{2}=1|h_{1}=1,r_0) = 0,\\
&P(h_{2}=1|h_{1}=0,r_0) = a.
\end{align}
We can use the probability relation
\begin{equation}
P(h_{2},h_{1}|r_0) =P(h_{2}|h_{1},r_0)P(h_{1}|r_0)
\end{equation}
to compute the two-searcher conditional probabilities
\begin{align}
&P(h_{2}=1,h_{1}=1|r_0) = 0,\\
&P(h_{2}=1,h_{1}=0|r_0) = a(1-a).
\end{align}
The other probability distributions
are computed using similar reasoning. The results are summarized in
Table~\ref{tab:1d}.
\begin{table}[h] 
\begin{center} 
\begin{tabular}{c|c|c|c|c|} 
    \hline    
    Cases&$h_{1},h_{2}$&$P(h_{1}|r_{0})$ &$P(h_{2}|r_{0}) $
    &$P(\{h_{1},h_{2}\}|r_{0})$\\
    \hline
    \hline 
                 &$ 1,1 $&$ a   $&$ a(1-a)   $&$ 0         $\\ 
    $ s_0s_1s_2 $&$ 1,0 $&$ a   $&$ 1-a(1-a) $&$ a         $\\ 
    $ s_2s_1s_0 $&$ 0,1 $&$ 1-a $&$ a(1-a)   $&$ a(1-a)    $\\ 
                 &$ 0,0 $&$ 1-a $&$ 1-a(1-a) $&$ (1-a)^{2} $\\ 
    \hline 
    \hline 
                 &$ 1,1 $&$ a   $&$ a        $&$ a^{2}     $\\ 
    $ s_1s_0s_2 $&$ 1,0 $&$ a   $&$ 1-a      $&$ a(1-a)    $\\ 
    $ s_2s_0s_1 $&$ 0,1 $&$ 1-a $&$ a        $&$ a(1-a)    $\\ 
                 &$ 0,0 $&$ 1-a $&$ 1-a      $&$ (1-a)^{2} $\\ 
    \hline
    \hline 
                 &$ 1,1$ &$ a(1-a)   $&$ a    $&$ 0         $\\ 
    $ s_0s_2s_1 $&$ 1,0$ &$ a(1-a)   $&$ 1-a  $&$ a(1-a)    $\\ 
    $ s_1s_2s_0 $&$ 0,1$ &$ 1-a(1-a) $&$ a    $&$ a         $\\
                 &$ 0,0$ &$ 1-a(1-a) $&$ 1-a  $&$ (1-a)^{2} $\\ 
    \hline 
\end{tabular}
\end{center} 
\caption{Conditional probabilities for the six possible arrangements of
the source $s_0$ and searchers $s_1,s_2$ shown in Fig.~\ref{fig:1d_diagram}.
Since the searchers are identical, $s_1$ and $s_2$ are interchangeable and
there are only three unique cases.
 In cases $s_0s_1s_2$ and $s_2s_1s_0$,
searcher one is between the source and searcher two. A particle emitted by the source in the direction of the searchers will reach searcher one first. If searcher one detects the particle, searcher two will not be able to detect it. Searcher two will only have a chance of detecting the particle if the particle passes through searcher one undetected.  Similarly, in cases 
 $s_0s_2s_1$ and $s_1s_2s_0$, the source is between the searchers and the searchers do not interfere with each other. Furthermore, since the source emits
two particles simultaneously in opposite directions, in only these cases do both searchers have a chance of each detecting a particle.} 
\label{tab:1d}  
\end{table}

Using Table~\ref{tab:1d}, we can analytically compute the
information theoretic quantities we need to determine synergy and
redundancy.  These are 
\begin{equation} 
R(h_1,h_2,r_0)=I(h_1,h_2)-I(h_1,h_2|r_0),
\label{eq:R}
\end{equation}
% \begin{equation} I(h_{1},h_{2}|r_{0})\equiv \sum_{h_{1},h_{2}}\int_{0}^{1}
%     dr_{0} P(h_{1},h_{2},r_{0}) \log_{2}
%     \frac{P(h_{1},h_{2}|r_{0})}{P(h_{1}|r_{0})P(h_{2}|r_{0})},
% \end{equation}
\begin{equation} I(h_{1},h_{2}) \equiv 
  \sum_{h_{1},h_{2}}
  P(h_{1},h_{2}) \log_{2} \frac{P(h_{1},h_{2})}{P(h_{1})P(h_{2})},
\end{equation} 
and 
\begin{eqnarray}
I(h_{1},h_{2}|r_{0}) &\equiv \sum_{h_{1},h_{2}}\int_{0}^{1}&
    dr_{0} P(h_{1},h_{2},r_{0})\\
&&    \times\log_{2}    \frac{P(h_{1},h_{2}|r_{0})}{P(h_{1}|r_{0})P(h_{2}|r_{0})};\nonumber
\end{eqnarray}
where the probability distributions are calculated
    as follows: 
    \begin{align} P(h_{1},h_{2},r_{0}) &= P(h_{1},h_{2}|r_{0})
        P(r_{0}),\\ P(h_{1},h_{2}) &= \int_{0}^{1} dr_{0} P(h_{1},h_{2}|r_{0})
        P(r_{0}),\\ P(h_{1}) &= \int_{0}^{1} dr_{0} P(h_{1}|r_{0}) P(r_{0}),\\
        P(h_{2}) &= \int_{0}^{1} dr_{0} P(h_{2}|r_{0}) P(r_{0}). \label{eq:h2int}
    \end{align}
Initially we consider a uniform probability distribution (prior) of source locations: $P(r_{0})=1$.  
% When integrating over $r_{0}$ to compute probability distributions such as in
% Eq.~\eqref{eq:h2int}, one must take into account which of the three cases are
% relevant. Consider a generic probability 
% distribution $f=f(r_{0},r_{1},r_{2},h_{1},h_{2})$ and let a subscript on $f$ denote the 
% probability distribution from the corresponding case (the subscript $012$ indicates
% the case $s_0s_1s_2$ and so on).  Regarding 
% Eq.~\eqref{eq:h2int}, $f = P(h_{2}|r_{0}) P(r_{0})$; the quantities for the 
% other integrals are similar.  Thus for $r_{1}<r_{2}$,
% \begin{equation} 
%     \int_{0}^{1} f dr_{0} =
%     \int_{0}^{r_{1}} f_{012} dr_{0} + \int_{r_{1}}^{r_{2}} f_{102} dr_{0} +
%     \int_{r_{2}}^{1} f_{120} dr_{0}.  \label{eq:flimits} 
% \end{equation}
% Similarly, for $r_{1}>r_{2}$: 
% \begin{equation} 
%     \int_{0}^{1} f dr_{0} = \int_{0}^{r_{1}} f_{021} dr_{0} + 
%     \int_{r_{1}}^{r_{2}} f_{201} dr_{0} + \int_{r_{2}}^{1} f_{210} dr_{0}.  
% \end{equation}
% By symmetry, for this example it is sufficient to consider only $r_{1}<r_{2}$.  

For small capture probability, $a$,
we can expand Eq.~(\ref{eq:R}) as a Taylor series in $a$ to
get an analytical solution for the critical values $a_{c}$ where $R|_{a=a_{c}}=0$.
For $r_{1}>r_{2}$, the critical values are given by
\begin{align}
    a_{c} &=  \sqrt{\frac{ 3(r_{2}-r_{1}) \log{(r_{1}-r_{2})}}
    {r_{1}^{3} - r_{2}^{3} - 3 r_{1}^{2} +  2 r_{1} +r_{2}}}.
    \label{eq:a_critical}
\end{align} 
For $r_{1}<r_{2}$, $a_c$ is given by Eq.~\eqref{eq:a_critical} under the transformations
$r_1 \to r_2$ and $r_2 \to r_1$.  These relations give the approximate
boundary between the regions of synergy and redundancy.  In the limit
$r_{1} \to r_{2}$, $R \to 0$.  For these formulas we
used a uniform source distribution $P(r_0)=1$ but it is possible to
repeat these calculations with a different $P(r_{0})$.

\begin{figure}[htb] 
  \includegraphics[width=\columnwidth]{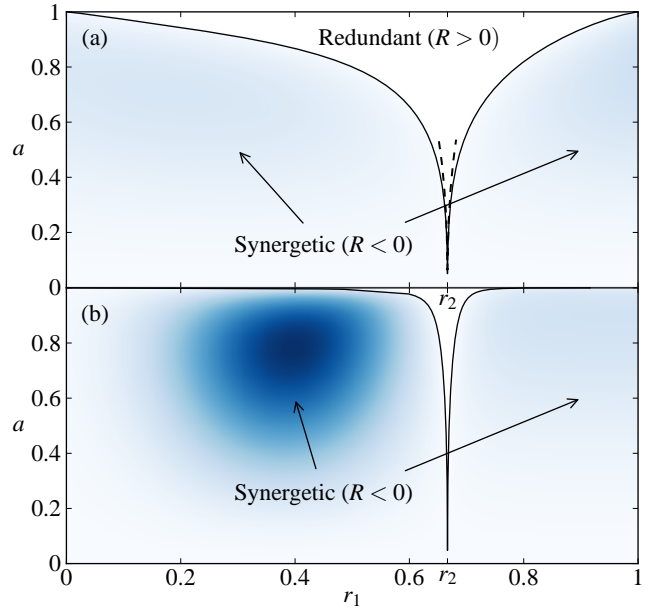}
     \caption{
       Synergetic and redundant regions for two searchers
       in the one-dimensional correlated search problem. 
       The value of $R(h_{1},h_{2},r_{0})$ is shown for different
       locations $r_1$ of $s_1$ with $s_2$ held fixed at $r_2=2/3$
       and the capture probability varying from $a=0$ (no capture) 
       to $a=1$ (complete capture).  Darker regions represent higher
       synergy (larger negative values of $R$) and the white areas
       represent redundancy.
       The critical contour line $R=0$ separates the synergetic and
       redundant regions.
       (a) When the source is equally likely to be anywhere, 
       $P(r_{0})=1$, there is only weak synergy.
       (b) Then the probability of the source is a Gaussian distribution 
       centered at $1/3$, 
       $P(r_{0}) \propto \exp(-(r_{0}-1/3)^{2})$, 
       the synergy is highest near the source at $1/3$ since the
       mutual information is highest there as well.
       The dashed line in (a) shows the critical values $a_c$ computed
       by the approximation in 
       Eq.~\eqref{eq:a_critical}; this
       approximation is good for $a_c$ {\scriptsize $\lesssim$} $0.3$.
}
  \label{fig:1dR}
\end{figure}

For larger values of $a$, when the expansion is no longer valid
the condition $R=0$ [as in Eq.~(\ref{eq:R})] can be solved numerically.  
Figure~\ref{fig:1dR} shows
$R(h_{1},h_{2},r_{0})$ as a function of the capture probability 
$a$ and searcher $s_1$ location $r_{1}$ 
with searcher $s_2$ fixed at  $r_{2}=2/3$.
This figure also shows $R(h_{1},h_{2},r_{0})$ for a Gaussian 
probability of the source location
$P(r_{0}) = B \exp(-(r_{0}-1/3)^{2})$, where $B$ is a normalization
factor. In both cases, $R>0$  (indicating redundancy) when
the searchers are close together and $R<0$ (indicating synergy)
when they are further apart.  For the Gaussian distribution source
location synergy is strongest when $s_1$ is close to the source because
the mutual information between $r_{0}$ and $h_{1}$ is peaked there as
well.

Figure~\ref{fig:1dR}(a) shows how the capture probability $a$
influences the possible strategies of the searchers.  If $a$ is close
to $1$, then the searchers only realize synergy if they are far apart,
which maximizes the chance of the source being between them.  But if
$a$ is is close to $0$ then nearly any arrangement of searchers is
synergetic but only weakly.  Figure~\ref{fig:1dR}(b) demonstrates the
significance of the prior $P(r_0)$ in the calculation of
$R(h_{1},h_{2},r_{0})$.  The variance of $P(r_0)$ is equal to $1$, the
size of the domain, making the Gaussian distribution quite broad.
Nonetheless, the fact that the prior is weakly peaked at some point in
the domain dramatically reduces the area of the redundant ($R>0$)
region and allows for much greater synergy between the searchers.
Although search problems in the real world are rarely one-dimensional,
this example illustrates the basic calculations for determining
synergy.

\section{Searching for correlated signals in two dimensions} 
The  one-dimensional example provides a tractable starting point
for generalization to two-dimensional problems.  The simplest generalization
is the case of two mobile searchers and one stationary source in a 
two-dimensional finite domain.  Imagine a source in which chemical or nuclear 
reactions are occurring and the products of the reactions leave the source 
with relative angular correlations.  
Our two-dimensional idealized problem consists of a source that
emits $2$ particles simultaneously at each time step in opposite
directions along some emission axis.  At each time step, a new
emission axis angle is chosen uniformly at random from $(0,2\pi]$. 
The particles move along straight-line trajectories but the searchers
are not able to measure the velocities of any detected particles, so
geometric methods such as triangulation are not possible.
\begin{figure}[htb]
  \centering
  \includegraphics[width=\columnwidth]{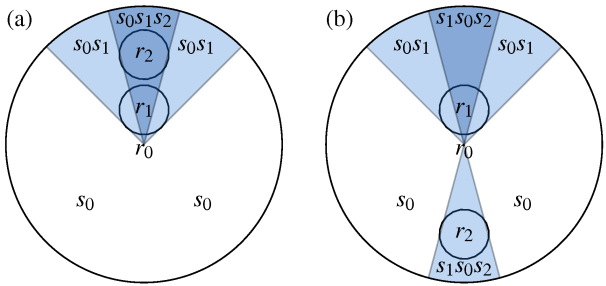} 
  \caption{ Diagram for the two-dimensional example.  The source position $r_0$
    is fixed in the center and the searchers are disks of radius $d$. At each
    time step the source emits two particles in in opposite directions
    with a random angle.   Each 
    possible emission axis passing through the source corresponds to one of six
    cases for source and searcher configurations as 
    given in Tables.~\ref{tab:1d} and \ref{tab:2d}.}
  \label{fig:2d_diagram}
\end{figure}

In the real world, searchers could be, for example, autonomous mobile
sensors capable of detecting radiation or other reaction products from
the above example.  Most modern-day autonomous robots do not move much
faster than a walking pace~\cite{Playter_06}, largely due to the
difficulties of navigating uncertain terrain safely.  Thus movement is
relatively costly (backtracking will take a significant amount of
time) and the searchers should make decisions to refine their
trajectories often, using any new available information.  For this
problem, this means that the searchers can only make small discrete
movements between measurements.

To represent simplified
mobile autonomous robots, we cast the searchers as identical disks of radius 
$d$ that move on a regular Cartesian grid.  Unlike the one-dimensional case
these searchers have spatial extent; there are two parts to the calculation of particle detection.
First, if a particle travels along a straight line trajectory
that passes through a searcher, the capture probability is $0<a<1$.  As in the 
one-dimensional example, if a particle is observed by a searcher, it is absorbed
so it cannot be observed by the other searcher.  Second, we must consider the 
probability that the particle's trajectory will pass through the searcher.  
This is a function of the searcher radius $d$ and the distance to the source.  
The variables $r_{0}$, $r_{1}$, and $r_{2}$ for the positions of 
the source and searchers each have two 
components, e.g.$(r_{0,x},r_{0,y})$, since they represent positions on
a two-dimensional grid.

As in the one-dimensional example, there are different cases for 
the probabilities which depend on the 
relative position of the two searchers to the source.
For all possible straight line emission axes
passing through the source, some lines may pass through no searchers (case $s_0$),
only searcher $s_1$ (case $s_0s_1$), or only searcher $s_2$ (case $s_0s_2$).  If a line passes
through both searchers, the source is between the searchers (case $s_1s_0s_2$ as
in the one-dimensional example), or one of the other searchers is in front of the other
(cases $s_0s_1s_2$ and $s_0s_2s_1$ as in the one-dimensional example).  These cases 
are illustrated in Fig.~\ref{fig:2d_diagram}.  For any source location, there will
be a range of angles $\Delta \theta_{c}$ for each case $c$.  The probabilities
for the cases that do not appear in the one-dimensional example (see Table~\ref{tab:1d})
are detailed in Table~\ref{tab:2d}.  Quantities such as $P(h_{1},h_{2}|r_{0})$
are a superposition of the values for the different cases, weighted by the
proportion of angles corresponding to each case
\begin{equation} 
    P(h_{1},h_{2}|r_{0}) = \sum_{c \in \{\text{cases}\}} 
    \frac{\Delta \theta_{c}}{2\pi}P_{c}(h_{1},h_{2}|r_{0}),
    %f = \sum_{c = \bigl\{\substack{0, 01, 02,\\
    %012, 021, 102}\bigr\}} \frac{\Delta \theta_{c}}{2\pi}f_{c}.
    \label{eq:2dfsum} 
\end{equation} 
and similarly for other quantities such as $P(h_{1}|r_{0})$, etc.  
While it is in principle possible to find the $\Delta\theta_{c}$ analytically
using geometry, in practice it is much more efficient to do this numerically.
\begin{table}[htb]
    \vspace{0.2cm}
  \centering
    \begin{tabular}{c|c|c|c|c|} 
    \hline 
    Case&$h_{1},h_{2}$&$P(h_{1}|r_{0})$ &$P(h_{2}|r_{0})$ &$P(\{h_{1},h_{2}\}|r_{0})$\\ \hline 
    \hline 
         &$ 1,1$ &$ 0  $&$ 0  $&$ 0 $\\ 
    $s_0$&$ 1,0$ &$ 0  $&$ 1  $&$ 0 $\\ 
         &$ 0,1$ &$ 1  $&$ 0  $&$ 0 $\\ 
         &$ 0,0$ &$ 1  $&$ 1  $&$ 1 $\\ 
    \hline 
    \hline
            &$ 1,1$ &$ a  $&$ 0  $&$ 0 $\\ 
    $s_0s_1$&$ 1,0$ &$ a  $&$ 1  $&$ a $\\ 
            &$ 0,1$ &$ 1-a  $&$ 0  $&$ 0 $\\ 
            & $ 0,0$ &$ 1-a  $&$ 1  $&$ 1-a $\\ 
    \hline 
    \hline
            &$ 1,1$ &$ 0  $&$ a $&$ 0 $\\ 
    $s_0s_2$&$ 1,0$ &$ 0  $&$ 1-a  $&$ 0 $\\
            &$ 0,1$ &$ 1  $&$ a  $&$ a $\\ 
            &$ 0,0$ &$ 1  $&$  1-a $&$ 1-a $\\ 
    \hline 
\end{tabular} 
    \caption{Conditional probabilities for cases unique to the two
      dimensional example in Fig.~\ref{fig:2d_diagram}.  The cases
      $s_1s_0s_2$, $s_0s_1s_2$, and $s_0s_2s_1$ (not shown),
      are identical to those of the one dimensional problem 
      given in Table~\ref{tab:1d}.  
      As shown in Figure 3, the case $s_0$ corresponds to the range of angles for which it is impossible for either searcher to detect a particle. Case $s_{0}s_{1}$ corresponds to the range of angles for which only searcher one has a chance of detecting a particle, and case $s_{0}s_{1}$ corresponds to the range of angles for which only searcher two has a chance of detecting a particle.}
   \label{tab:2d}%
  \end{table}

Note that the different cases in the one-dimensional example were taken into 
account implicitly during integration as in Eq.~\eqref{eq:h2int}.
In the two-dimensional example they are taken into account
when the effective probability distributions are computed as a superposition
of the probabilities of the individual cases.  This is because there is more
than one case for each source location relative to the searchers.

We illustrate the two-dimensional searcher problem with a setup
of searcher and source locations that illustrate the synergetic
and redundant positions. We use the Gaussian prior for the
initial guess 
\begin{equation}
    P(r_{0}) = A\exp{-\frac{||r_{0}-s||^{2}}{\sigma^{2}}},
\end{equation}
where $A$ is the normalization and $\sigma$ determines the overall shape of the
distribution.  The vector $s=(2,2)$ is the most probable position of the source.
The quantity $R(h_{1},h_{2},r_{0})$ [see Eq.~\eqref{eq:Rdefn}]
determines whether the searchers are positioned 
synergetically relative to the source.  When the searchers are performing an 
infotaxis search, realizing synergetic relative positions will in principle lead
to the fastest reduction in uncertainty.  

\begin{figure}[htb] 
  \begin{center} 
    \includegraphics[width=\linewidth]{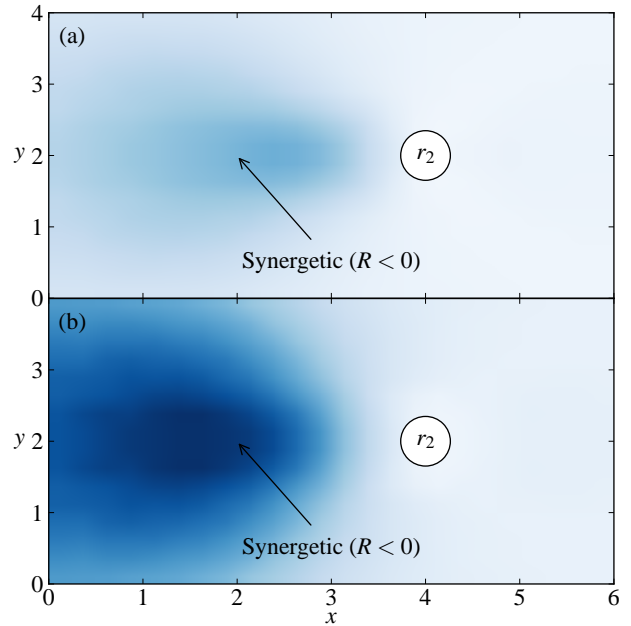}
    \caption{Synergy for two searchers for two-dimensional
      correlated signals.  The value of $R(h_{1},h_{2},r_{0})$ 
      is shown as a function of the position $r_{1}$ of searcher $s_1$
      for two different values of the capture probability $a$: 
      (a) $a=0.25$; (b) $a=0.75$.  The position of searcher $s_2$ 
      is fixed $r_{2}=(2,4)$.  The most probable source location 
      (the peak of the Gaussian distribution for the initial source 
      location) is at $r_0=(2,2)$.  Darker blue corresponds to stronger
      synergy ($R<0$).  Synergy is strongest for large $a$ and when 
      searcher $s_1$ is near the source and not behind searcher $s_2$ 
      relative to the source.}
  \label{fig:2dR}
\end{center} 
\end{figure}

In Fig.~\ref{fig:2dR} we plot $R$ as a function of the position of one
searcher $r_{1}$ with the position of the second searcher $r_{2}$ held
fixed.  We find for this example that only synergy ($R<0$) is possible.  
The light areas in Fig.~\ref{fig:2dR} correspond to weak
synergy and the dark areas to stronger synergy.
As in the one-dimensional example
synergy is strongest for large
$a$ and when searcher $s_1$ is near the source and not behind searcher
$s_2$.  A small cross section
[$a=0.25$ in Fig.~\ref{fig:2dR}(a)] allows only for weak synergy,
whereas a larger cross section [$a=0.75$ in Fig.~\ref{fig:2dR}(b)]
gives much stronger synergy for certain relative positions.    The
strongest synergy comes from a large cross section paired with optimal
positioning.  The maximum synergy is realized
when searcher $s_1$ is close to the peak of $P$
and especially when the peak of $P$ is between the searchers,
Note that unfavorable positions (such as $r_1 = (5,2)$, behind searcher
$s_2$) provide minimal synergy regardless of the value of $a$.

For real-world problems, larger cross sections yield more information
and therefore stronger synergy is possible.  However for some
applications the cross section of a sensor on a searcher may be
limited by practical considerations such as weight or power
consumption.  This example shows that synergy is still possible for
small $a$.  Furthermore, this example emphasizes the importance of the
probability estimate of source locations $P(r_0)$.  The infotaxis
algorithm is designed such that the searchers will explore, gathering
new information, if their arrangement is not sufficiently synergetic
to warrant exploitation.  This allows the searchers to succeed even
with no starting information, but they may not realize strong synergy
until the estimate of $P(r_0)$ is sufficiently refined.

\section{Conclusion} 

In this work we studied search algorithms for autonomous agents looking for the spatial location of a stochastic source. In spatial search problems, since the exploitation of synergy requires spatial or temporal correlations, we considered problems in which a source emits two particles simultaneously in opposite directions. This is a simplification of physical problems in which there is a reaction and the products travel in directions that have angular correlations. We showed that both synergy and redundancy are possible for one-dimensional search problems but not for two-dimensional searches, where only synergy is possible. Since even unfavorable arrangements of searchers are synergetic, in two-dimensional search problems like these coordination is always advantageous.

Simple examples such as these that can be studied analytically provide insight into real-world problems. In real-world problems, there will necessarily be additional considerations. It may not always be possible to write a closed-form equation for the nature of the correlations in the signal from the source. It may be necessary to directly measure any correlations and use this to estimate capture probabilities. Furthermore, various probabilities may not be stationary in time or the signals from the source may get progressively weaker. For example, signals from a radio transmitter may decrease in strength over time as its batteries are slowly exhausted. An additional consideration is that in the real-world, communication between agents may only be possible at certain times and may not be instantaneous as in our simple examples. These general considerations are crucial for the exploitation of multi-agent infotaxis in terms of the design of optimal collective algorithms in particular applications. The next steps for making this approach applicable to a broader class of problems, including those not limited to spatial searches\cite{Bettencourt_2009}, are to generalize the results to more
than two searchers and to explore how synergy may be best leveraged to
give increases in search speed and efficiency.

\begin{acknowledgements}
This work was supported in part by a DCI IC Postdoctoral Fellowship and the Department of Energy at Los Alamos National Laboratory under contract 
DE-AC52-06NA25396 through the Laboratory Directed Research and Development Program.  This is published under release LA-UR 09-07676.

\end{acknowledgements}

%\bibliographystyle{plain}
%\bibliography{synergy}

\begin{thebibliography}{10}

\bibitem{Agogino_2008}
A.~K. Agogino and K.~Tumer.
\newblock Analyzing and visualizing multiagent rewards in dynamic and
  stochastic domains.
\newblock {\em Autonomous Agents and Multi-Agent Systems}, 17(2):320, 2008.

\bibitem{Bernardo_1994}
J.~M. Bernardo and A.~F.~M. Smith.
\newblock {\em Bayesian Theegesory}.
\newblock Wiley, New York, 1994.

\bibitem{Bettencourt_2009}
L.~M.~A. Bettencourt.
\newblock The rules of information aggregation and emergence of collective
  intelligent behavior.
\newblock {\em Topics in Cognitive Science}, 1:598, 2009.

\bibitem{Bettencourt_2008}
L.~M.~A. Bettencourt, V.~Gintautas, and M.~I. Ham.
\newblock Identification of functional information subgraphs in complex
  networks.
\newblock {\em Phys. Rev. Lett.}, 100:238701, 2008.

\bibitem{Bettencourt_2007}
L.~M.~A. Bettencourt, G.~J. Stephens, M.~I. Ham, and G.~W. Gross.
\newblock Functional structure of cortical neuronal networks grown in vitro.
\newblock {\em Phys. Rev. E}, 75:021915, 2007.

\bibitem{Cover_1991}
T.~M. Cover and J.~A. Thomas.
\newblock {\em Elements of Information Theory}.
\newblock Wiley, New York, 1991.

\bibitem{Eisenbach_2004}
M.~Eisenbach and J.~W. Lengeler.
\newblock {\em Chemotaxis}.
\newblock Imperial College Press, London, 2004.

\bibitem{Fine_2002}
S.~Fine, R.~Gilad-Bachrach, and E.~Shamir.
\newblock Query by committee, linear separation and random walks.
\newblock {\em Theor. Comput. Sci.}, 284(1):25, 2002.

\bibitem{Fox_2001}
J.~Fox, D.~Glasspool, and J.~Bury.
\newblock Quantitative and qualitative approaches to reasoning under
  uncertainty in medical decision making.
\newblock In {\em Artificial Intelligence in Medicine: Lecture Notes in
  Computer Science}, volume 2101, pages 272--282. Springer, Berlin/Heidelberg,
  Germany, 2001.

\bibitem{Freund_1997}
Y.~Freund, E.~Shamir, and N.~Tishby.
\newblock Selective sampling using the query by committee algorithm.
\newblock In {\em Machine Learning}, page 133, 1997.

\bibitem{Gintautas_SBP09}
V.~Gintautas, A.~Hagberg, and L.~M.~A. Bettencourt.
\newblock When is social computation better than the sum of its parts?
\newblock In H.~Liu, J.~J. Salerno, and M.~J. Young, editors, {\em Social
  Computing, Behavior Modeling, and Prediction}, page~93, 2009.

\bibitem{Hollinger_2009}
G.~Hollinger, S.~Singh, J.~Djugash, and A.~Kehagias.
\newblock Efficient multi-robot search for a moving target.
\newblock {\em The International Journal of Robotics Research}, 28(2):201--219,
  2009.

\bibitem{Playter_06}
R.~Playter, M.~Buehler, and M.~Raibert.
\newblock Bigdog.
\newblock In G.~R. Gerhart, C.~M. Shoemaker, and D.~W. Gage, editors, {\em
  Proc. SPIE: Unmanned Systems Technology {VIII}}, volume 6230, 2006.

\bibitem{Russell_2003}
R.~A. Russell, A.~Bab-Hadiashar, R.~L. Shepherd, and G.~G. Wallace.
\newblock A comparison of reactive robot chemotaxis algorithms.
\newblock {\em Robotics and Autonomous Systems}, 45(2):83, 2003.

\bibitem{Schneidman_2003}
E.~Schneidman, W.~Bialek, and M.~J. {Berry II}.
\newblock Synergy, redundancy, and independence in population codes.
\newblock {\em J. Neurosci.}, 23:11539, 2003.

\bibitem{Seung_1992}
H.~S. Seung, M.~Opper, and H.~Sompolinsky.
\newblock Query by committee.
\newblock In {\em COLT '92: Proceedings of the fifth annual workshop on
  Computational learning theory}, page 287, 1992.

\bibitem{Sulton_1998}
R.~S. Sulton and A.~G. Barto.
\newblock {\em Reinforcement learning: an introduction}.
\newblock MIT Press, Cambrigde MA, 1998.

\bibitem{Vergassola_2007}
M.~Vergassola, E.~Villermaux, and B.~I. Shraiman.
\newblock ``{I}nfotaxis'' as a strategy for searching without gradients.
\newblock {\em Nature}, 445:406, 2007.

\bibitem{Zhang_2002}
W.~Zhang, Z.~Deng, G.~Wang, L.~Wittenburg, and Z.~Xing.
\newblock Distributed problem solving in sensor networks.
\newblock In {\em AAMAS '02: Proceedings of the first international joint
  conference on Autonomous agents and multiagent systems}, pages 988--989, New
  York, NY, USA, 2002. ACM.

\end{thebibliography}

\end{document}